\documentclass[amsmath, amsfonts, superscriptaddress, twocolumn, prb]{revtex4}
\usepackage{graphicx}
\usepackage{epsfig}
\usepackage{bm}
\usepackage{dcolumn}
\usepackage{amsmath}
\usepackage{amssymb}
\usepackage{euscript}

\def\rr{\bm{r}}
\def\pp{\bm{p}}
\def\qq{\bm{q}}

\begin{document}

\title{Density of states and magnetotransport in Weyl semimetals with long-range disorder}

\author{D. A. Pesin}
\affiliation{Department of Physics and Astronomy, University of Utah, Salt Lake City, UT 84112, USA}

\author{E. G. Mishchenko}
\affiliation{Department of Physics and Astronomy, University of Utah, Salt Lake City, UT 84112, USA}

\author{A. Levchenko}
\affiliation{Department of Physics, University of Wisconsin-Madison, Madison, Wisconsin 53706, USA}
\affiliation{Institut fur Nanotechnologie, Karlsruhe Institute of Technology, 76021 Karlsruhe, Germany}

\begin{abstract}
We study the density of states and magnetotransport properties of disordered Weyl semimetals, focusing on the case of a strong long-range disorder. To calculate the disorder-averaged density of states close to nodal points, we treat exactly the long-range random potential fluctuations produced by charged impurities, while the short-range component of disorder potential is included systematically and controllably with the help of a diagram technique. We find that for energies close to the degeneracy point, long-range potential fluctuations lead to a  finite density of states. In the context of transport, we discuss that a self-consistent theory of screening in magnetic field may conceivably lead to non-monotonic low-field magnetoresistance.
\end{abstract}

\date{\today}
\maketitle

\section{Introduction}

Coulomb disorder plays dominant role in determining thermodynamic and kinetic properties of doped semiconductors due to its long-range nature.~\cite{SE-Book} This physics becomes especially intriguing in the context of Dirac materials, namely solid state systems that host quasiparticles with linear dispersion near the band-touching degeneracy points, such as that in $d$-wave superconductors, \cite{BalatskyReview} graphene, \cite{Graphene-Review-1,Graphene-Review-2} topological insulators \cite{TI-Review-1,TI-Review-2} and Weyl semimetals. \cite{Turner2013,Vafek,Burkov-Review} While the former three classes of systems have been intensively studied over almost a decade by now, materials that can be identified as Weyl semimetals have been discovered only fairly recently. \cite{Kim2013,WSM-Exp-1,WSM-Exp-2,WSM-Exp-3,WSM-Exp-4,WSM-Exp-5,WSM-Exp-6,WSM-Exp-7,WSM-Exp-8}

Theoretical studies of the effect of disorder on a single Weyl node have a long history, \cite{Fradkin} and this problem has been revisited in a number of recent works. \cite{Herbut, Ominato-1,Sbierski,Syzranov-1,Skinner,Altland} The prevailing point of view is that a weak disorder has negligible effect on the density of states, which vanishes quadraticaly with the energy counted from the nodal point. This behavior persists up to a certain critical disorder strength beyond which the density of states acquires a finite value at zero energy. Obviously, such scenario should have serious implications for the transport properties of WSMs that rely on their semimetallic nature (a vanishing density of states at the nodal points). \cite{Burkov-Hook,Hosur,Ominato-2,Syzranov-2,Rodionov-1,Rodionov-2} At strong disorder singular transport features are eliminated. It is thus important to have a controlled theory of disorder effects on a single nodal point. Conventional treatments based on the self-consistent Born approximation, though can be qualitatively correct, are uncontrolled near Dirac or Weyl points. \cite{Sbierski}

Magnetotransport properties of WSMs are also expected to have a highly unusual character owing to the unconventional Landau quantization for the three-dimensional Dirac spectrum, and the multiband nature of Weyl semimetals.  In particular, a strong classical negative magnetoresistance was predicted for Weyl systems with long intravalley relaxation times, \cite{Son-Spivak,Spivak-Andreev} as well as robust linear positive magnetoresistance in the quantum limit, when a single Landau level is occupied. \cite{Abrikosov} Both types of behavior were recently observed in experiments \cite{Ong,Feng} on a candidate Dirac (``double-Weyl'') material Cd$_3$As$_2$, yet at this point it cannot be stated with confidence whether the experimental observations are in full correspondence with theoretical models. For various recent theoretical and experimental aspects of magnetotransport results in Dirac semimetals see Refs.~\onlinecite{Yang,Gorbar,Ghimire,Burkov-MR,Shen-1,Shen-2,Goswami,Adam,Klier,Shekhar}.

In light of the recent experimental advances that triggered a flood of theoretical work, our essential motivation is to investigate the role of Coulomb impurities on properties of WSMs. In contrast to the model of short-range disorder, which so far received most of theoretical attention, we obtain results that are controlled by a small parameter in the limit of \emph{strong} disorder. We also observe that due to the delicate interplay of Landau quantization and screening effects, the low-field magnetoresistance may display nonmonotonic behavior. The latter feature is highly sensitive to the strength of interaction and thus is not universal.

\section{Model of a Weyl semimetal with long-range disorder}

The quasiparticle dispersion in the vicinity of a single isotropic nodal point can be described by the effective Hamiltonian ( $\hbar=c=1$ throughout)
\begin{equation}\label{H}
H=v{\bm \sigma}\cdot \pp+u(\rr),
\end{equation}
where $v$ is the Fermi velocity, and ${\bm \sigma}$ and $\pp$ are vector of the Pauli matrices, and momentum measured relative to the nodal point. The total potential as seen by an electron at position $\rr$ due to randomly distributed charged impurities at positions $\rr_i$ is taken in the form of a Yukawa-like potential
\begin{equation}\label{u}
u(\rr)=\sum_i\phi(\rr-\rr_i),\quad \phi(\rr)=\frac{e^2}{\epsilon r}\exp(-\kappa r).
\end{equation}
Here index $i$ labels impurities, $\epsilon$ is the dielectric constant of the host material, and $\kappa$ is the inverse screening radius. We assume that impurity positions are completely random, and calculate the correlator of the disorder potential,
\begin{eqnarray}\label{eq:fulldisorder}
\langle u(\rr)u(\rr')\rangle=\left(\frac{4\pi e^2}{\epsilon}\right)^2\int\frac{d^3q}{(2\pi)^3}\frac{n\exp(i\qq(\rr-\rr'))}{(q^2+\kappa^2)^2}\nonumber\\
=\frac{2\pi ne^4}{\epsilon^2\kappa}\exp(-\kappa|\rr-\rr'|),\label{uu}
\end{eqnarray}
where $n$ is the concentration of impurities. As follows from Eq.~(\ref{uu}), the correlation length of the disorder potential is given by the screening radius, and the magnitude of its fluctuations is given by $\langle u^2(\rr)\rangle\equiv w^2=2\pi ne^4/\epsilon^2\kappa$. The Gaussian approximation is sufficient to describe the disorder potential fluctuations when there are many impurities in the screening volume, $n\kappa^{-3}\gg1$. The model of a disordered Weyl semimetal described above will be used in the forthcoming calculations. The validity of the Gaussian approximation will be checked \textit{a posteriori}, by considering the screening of Coulomb impurities by electrons moving in a self-consistently screened disorder potential, see Eq.~(\ref{nuandkappa}), and the text below it. A detailed account of self-consistent screening in the present problem is given in Ref. \onlinecite{Skinner}.

\section{Density of states at the nodal point}

In what follows we consider the density of states (DoS) close to the nodal point, $\varepsilon\to0$, where $\varepsilon$ is the electron energy.

DoS at the nodal point crucially depends on the relation between the strength of disorder, and the energy scale associated with its finite correlation length, $\kappa v$. For $w\ll \kappa v$, the disorder is weak, and the Born approximation can be used, with the expected result of zero DoS at the nodal point. For strong disorder, $w\gg \kappa v$, self-consistent Born approximation predicts a finite DoS at the nodal point,\cite{Ominato-1} but the approximation itself is uncontrolled. \cite{Sbierski}

To obtain systematic results, controlled by a small parameter, for the DoS and later for transport properties, we restructure the perturbation theory for the disorder potential similarly to the way it was done for heavily doped semiconductors. \cite{SE-Book, Efros} A crucial observation is that for $w\gg\kappa v$, the landscape of the disorder potential is smooth, in the sense that the typical length scale at which it changes, $1/\kappa$, is large compared to the typical electron wavelength $\lambda\sim v/w$. (The latter estimate assumes that it is the \emph{total} energy of the electron that is close to the nodal point, thus the kinetic energy is of the order of $w$.) This implies that the electron motion is semiclassical. Therefore, one may use the Thomas-Fermi approximation (rather than the clean system) as the ``reference point'' around which a perturbative expansion is developed.

To illustrate the use of the Thomas-Fermi (TF) approximation in this case, we consider the single particle Green's function for the Hamiltonian (\ref{H}):
\begin{equation}\label{G-u}
G^{R(A)}[u(\rr)]=\frac{1}{\varepsilon_\pm-v\bm{\sigma}\cdot\bm{p}-u(\rr)},
\end{equation}
where $\varepsilon_\pm=\varepsilon\pm i0$, and $u(\rr)$ plays a role of the effective local chemical potential. In general, since the electron momentum is an operator, one cannot simply average the Green's function over the distribution of $u(\rr)$. However, in the semiclassical limit we can treat $u(\rr)$ as a number within TF-approximation and thus obtain
\begin{equation}\label{G}
\langle G^{R(A)}\rangle=\int du \, F[u]\frac{1}{\varepsilon_\pm-v{\bm \sigma}\cdot\pp-u},
\end{equation}
where the distribution function for the local values of the disorder potential, $F[u]$, is defined according to
\begin{equation}
F[u]=\langle\delta(u-u(\rr))\rangle.
\end{equation}
Taking the imaginary part of the retarded component of $\langle G\rangle$ gives the density of states,
\begin{equation}\label{eq:oldDoS}
\nu(\varepsilon)=\int du F[u]\nu_0(\varepsilon-u),
\end{equation}
where $\nu_0(\varepsilon)$ is the DoS of the clean system. In the case of a Gaussian disorder potential defined in Eq.~(\ref{uu}), one obtains
\begin{equation}\label{P}
F[u]=\frac{1}{\sqrt{2\pi}w}\exp(-u^2/2w^2).
\end{equation}
Using $\nu_0(\varepsilon)=g\int\frac{d^3p}{(2\pi)^2}\delta(\varepsilon-vp)=g\varepsilon^2/2\pi^2v^3$, with $g$ being the degeneracy of the nodal point for a given Weyl system, and performing an elementary integration, one finds
\begin{equation}\label{nu}
\nu(\varepsilon)=\frac{g}{2\pi^2v^3}(\varepsilon^2+w^2),
\end{equation}
that is, a nonzero value for $\varepsilon=0$.

In the context of doped semiconductors, the semiclassical equation for the density of states, Eq.~(\ref{eq:oldDoS}), was suggested long ago, e.g. in Refs. \onlinecite{Bonch-Bruevich,Kane,Keldysh,Efros}, and has been recently used to describe DoS and nonlinear screening in Dirac systems in Ref. \onlinecite{Skinner}

\begin{figure}
  \centering
  \includegraphics[width=3.5in]{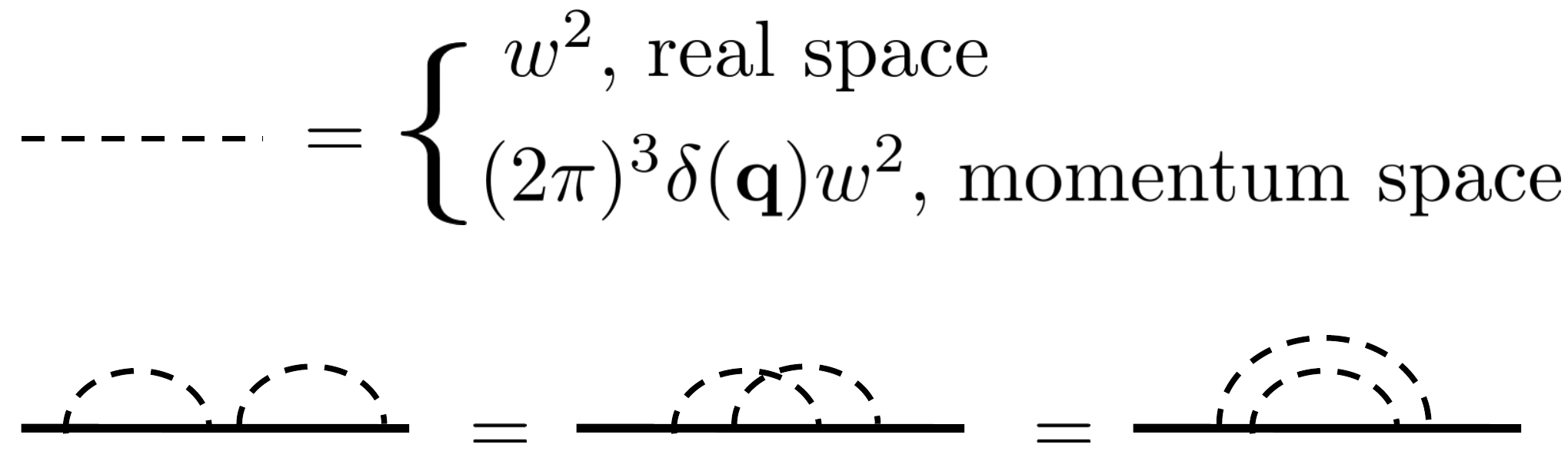}
  \caption{Diagrams that correspond to the second-order disorder correction to the Green's function. For the impurity line that does not depend on spatial coordinates, $\langle u(\rr)u(\rr')\rangle=w^2$, all the diagrams have the same value of $w^4 G_0^5$, where $G_0$ is the unperturbed Green's function.}\label{fig:diagrams}
\end{figure}

We motivated equation~(\ref{G}) on physical grounds, but it can be rigorously shown that it is exact in the $v\kappa/w\to0$ limit. One can formally send $\kappa\to0$ while keeping $w$ constant by considering $e^2\to0$, $n\to\infty$ limit with $en^{1/3}$ kept constant (see below). In this limit the disorder correlator is $\langle u(\rr)u(\rr')\rangle\approx w^2$, and describes disorder with \emph{infinite} correlation length. The problem of finding the single-particle Green's function for such a disorder potential (the so-called ``Keldysh model'') is exactly solvable, since the impurity lines in the standard technique do not transfer momentum, thus all diagrams to a given order have the same value, Fig.~\ref{fig:diagrams}. The full resummation of the perturbation theory becomes a combinatorial problem, solving which~\cite{Sadovskii} requires determining the number of ways $n$ impurity lines can be attached to $2n+1$ bare Green's functions in the $n^{\text{th}}$ order of perturbation theory (the second order diagrams are shown in Fig.~\ref{fig:diagrams}). The results for the disorder-averaged single-particle Green's functions and the DoS are precisely the above Eqs.~(\ref{G}) and~(\ref{eq:oldDoS}).

In what follows we consider the corrections to the result of Eq.~(\ref{nu}), following the treatment of Ref.~\onlinecite{Efros}.

\subsection{Improved perturbation theory for long-range disorder}

To build an improved perturbation theory for the potential defined by Eq.~(\ref{uu}), we formally rewrite the latter equation as
\begin{equation}\label{udecomposition}
  \langle u(\rr)u(\rr')\rangle=w^2+w^2\left(\exp(-\kappa|\rr-\rr'|)-1\right).
\end{equation}
If the second term in the right hand side of the above equation is neglected (the corresponding condition to be formulated below), one recovers the case of an infinite-range disorder potential, which was discussed above.

Going beyond the Keldysh model, when is it appropriate to treat the second term in the right hand side of Eq.~(\ref{udecomposition}) as a small correction? It is clear that one should require that $\kappa|\rr-\rr'|\ll1$ for the relevant values of $|\rr-\rr'|$. Since the solution for the semiclassical problem is represented by an ensemble average over samples having a random, but spatially uniform chemical potential of order $w$, see Eqs.~(\ref{G}) and (\ref{P}), the relevant value of $|\rr-\rr'|$ is of the order of the wavelength corresponding to the energy $w$, $|\rr-\rr'|\sim v/w$. Hence the corrections to the semiclassical approximations are expected to be small for $\kappa v/w\ll1$, coinciding with the condition for the validity of the semiclassical approach formulated above.

Our next immediate goal is to find the leading $\sim O(v\kappa/w)$ correction to Eq.~(\ref{nu}) due to the short-range component of disorder potential. As we are going to see, it is regular and free from divergencies near the Dirac point. The full disorder-averaged Green's function can be written as~\cite{Efros}
\begin{equation}\label{Gfull}
\langle G^{R(A)}\rangle=\int du \, F[u]\frac{1}{\varepsilon_\pm-u-v{\bm \sigma}\cdot\pp-\Sigma}.
\end{equation}
Here, for every value of $u$, the self-energy $\Sigma$ is determined from the usual diagram technique for disordered metals, in which the role of the bare Green's function is played by Eq.~(\ref{G-u}) with $u(\rr)$ replaced with $u$, and the impurity line (in momentum space), $D(\qq)$, is determined \emph{only} by the second term in the right hand side of Eq.~(\ref{udecomposition}):
\begin{equation}\label{impurityline}
  D(\qq)=\frac{8\pi w^2\kappa}{(q^2+\kappa^2)^2}-(2\pi)^3w^2\delta(\qq),\quad\int d^3q D(\qq)=0.
\end{equation}
In Eq.~(\ref{impurityline}) $\delta(x)$ is the Dirac delta-function. Its appearance signals that the long-range part of the impurity potential is subtracted from $D(\qq)$, as it has been already treated exactly by the averaging over the values of $u$. We will see below that after such removal, a simple Born approximation is valid for the short-range part of the disorder potential if $\kappa v\ll w$.

Nonetheless, below we set up the self-consistent Born approximation (SCBA) for the self-energy for the modified disorder correlator, Eq.~(\ref{impurityline}). We will not go beyond the first Born approximation in actual calculations, but having the full SCBA set up will allow us to contrast the present treatment with the conventional SCBA for long-range disorder.

Using the symmetry-suggested decomposition for the self-energy
\begin{equation}
\Sigma=\Sigma_0+\Sigma_1{\bm \sigma}\cdot\bm{n}_{\bm p},
\end{equation}
where $\bm{n}_{\bm p}$ is the unit vector in the direction of $\pp$, and introducing  $\xi_\pm=\varepsilon-u\pm i0$, for the retarded component we have
\begin{eqnarray}\label{SCBA-0}
&&\hskip-.25cm\Sigma_0(p)=\frac{2}{\pi}\int^{\infty}_{0}dqq^2\int^{+1}_{-1}dx\frac{\kappa w^2g_0(q)}{(p^2+q^2+\kappa^2-2pqx)^2}\nonumber\\
&&\hskip-.25cm-w^2g_0(p),\\
&&\hskip-.25cm\Sigma_1(p)=\frac{2}{\pi}\int^{\infty}_{0}dqq^2\int^{+1}_{-1}dx\frac{\kappa w^2xg_1(q)}{(p^2+q^2+\kappa^2-2pqx)^2}\nonumber\\
&&\hskip-.25cm-w^2g_1(p),\label{SCBA-1}
\end{eqnarray}
where we have introduced
\begin{eqnarray}
g_0=\frac{\xi_+-\Sigma_0}{(\xi_+-\Sigma_0)^2-(vp+\Sigma_1)^2},\\
g_1=\frac{vp+\Sigma_1}{(\xi_+-\Sigma_0)^2-(vp+\Sigma_1)^2}.
\end{eqnarray}

To simplify equations, we first introduce new functions $X_p=\xi_+-\Sigma_0$ and $Y_p=vp+\Sigma_1$. Second, we complete $x$-integration in Eqs.~(\ref{SCBA-0}) and (\ref{SCBA-1}) and observe that upon changing $q\to-q$ the corresponding integrands are even and odd functions respectively. We thus extend the $p$-integration to the whole axis such that $X_p$ is considered as an even function while $Y_p$ is odd. As a result we have to solve the following set of coupled integral equations
\begin{eqnarray}\label{eq:XY}
&&\hskip-.75cm
X_p=\xi_+-\frac{\kappa w^2}{\pi}\int^{+\infty}_{-\infty}
\frac{K_0(p,q)X_qdq}{X^2_q-Y^2_q}+w^2\frac{X_p}{X^2_p-Y^2_p},\nonumber\\
&&\hskip-.75cm
Y_p=vp+\frac{\kappa w^2}{\pi}\int^{+\infty}_{-\infty}\frac{K_1(p,q)Y_qdq}{X^2_q-Y^2_q}-w^2\frac{Y_p}{X^2_p-Y^2_p},
\end{eqnarray}
where the kernels are given explicitly by
\begin{equation}
\begin{split}
&K_0(p,q)=\frac{q}{p}\frac{1}{(p-q)^2+\kappa^2},\\
&K_1(p,q)=\frac{q}{p}\left[\frac{1}{(p-q)^2+\kappa^2}+\frac{1}{4pq}\ln\left(\frac{(p-q)^2+\kappa^2}{(p+q)^2+\kappa^2}\right)\right].
\end{split}
\end{equation}

\subsection{$\kappa\to0$ limit within the self-consistent Born approximation and Keldysh model}

It is instructive to understand how the limit of very long correlation length, $\kappa\to0, w=\text{const}$, is reached within various approximation schemes. Firstly, the results for the Keldysh model, Eqs.~(\ref{G}) and (\ref{eq:oldDoS}), are exact in this limit, as we have already discussed. In particular, this means that the quantities $X$ and $Y$, that determine the self-energy at the SCBA level for the modified disorder correlator, must attain bare values $X_p=\xi_+$, $Y_p=vp$ when Eqs.~(\ref{eq:XY}) are solved. That this is indeed the case can be seen noting that in the $\kappa\to0$ limit the kernels $K_{0,1}$ satisfy
\begin{equation}
\begin{split}
&\lim_{\kappa\to 0}\frac{\kappa}\pi K_{0,1}(p,q)=\delta(p-q).
\end{split}
\end{equation}
This implies that the second terms on the right hand sides of Eqs.~(\ref{eq:XY}) are exactly cancelled by the third terms. In fact, all self-energy corrections for the modified disorder correlator, not just the SCBA ones, vanish in the $\kappa\to0$ limit: This is the essence of the Keldysh model, in which the full diagrammatic series is summed up \emph{exactly} in the $\kappa\to 0$ limit, to yield Eq.~(\ref{G}) for the Green's function.~\cite{Efros,Sadovskii}

We can contrast the Keldysh model with the conventional SCBA, in which the full disorder correlator, Eq.~(\ref{eq:fulldisorder}), is used as the impurity line. This approximation can easily be obtained from Eqs.~(\ref{eq:XY}) by setting $\xi\equiv\epsilon-u\to\epsilon$, and dropping the third terms on the right hand sides of the equations for $X$ and $Y$. To distinguish this case from the rest of the paper, we denote $X$ and $Y$ for SCBA with full disorder correlator as $X^{B}$ and $Y^{B}$. To obtain simple analytic results, we limit ourselves to $\epsilon=0$. In the $\kappa\to 0$ limit the SCBA equations are
\begin{eqnarray}
&&\hskip-.75cm
X^B_p=i\delta-w^2
\frac{X^B_p}{(X^B_p)^2-(Y^B_p)^2},\nonumber\\
&&\hskip-.75cm
Y^B_p=vp+w^2\frac{Y^B_p}{(X^B_p)^2-(Y^B_p)^2}.
\end{eqnarray}
The infinitesimal $\delta$ is kept as a reminder that $\Im X_p>0$. $X^B_p$ and $Y^B_p$ can be easily found noting that the former is purely imaginary, while the latter is real:
\begin{eqnarray}\label{eq:solutionXY}
&&\hskip-.75cm
X^B_p=i\theta(2w-vp)\sqrt{w^2-\frac{(v p)^2}{4}},\nonumber\\
&&\hskip-.75cm
Y^B_p=\frac{vp}2+\theta(vp-2w)\sqrt{\frac{(v p)^2}{4}-w^2}.
\end{eqnarray}

The corresponding DoS can be found as
\begin{equation}
  \nu^B(\epsilon=0)=-\frac{2}\pi\Im\int\frac{d^3p}{(2\pi)^3}\frac{X^B_p}{(X^B_p)^2-(Y^B_p)^2},
\end{equation}
which upon using Eqs.~(\ref{eq:solutionXY}) gives
\begin{equation}\label{eq:nuBorn}
  \nu^B(\epsilon=0)=\frac{w^2}{2\pi^2 v^3}.
\end{equation}
In this treatment inter-valley scattering is neglected, and the result Eq.~(\ref{eq:nuBorn}) should be understood as the DoS per node.

We see that the SCBA with the full disorder correlator reproduces the exact answer for DoS in $\kappa\to0$ limit. This means that if one naively tries to correct SCBA result for DoS by including the low-order interference corrections to the self-energy, \cite{Ominato-2} the result will become worse, not better.

What the conventional SCBA does not capture is the form of the spectral weight that results from summation of the full diagrammatic series in the $\kappa\to0$ limit. As a function of momentum for a given energy $\epsilon$, the spectral weight for the Green's function of Eq.~(\ref{G}) is a Gaussian peak centered at $|\epsilon/v|$, having a width of $~w/v$. This exact result is clearly not reproduced by the conventional SCBA at low energies, Eq.~(\ref{eq:solutionXY}). For illustration purposes, we plot the total spectral weight (for conduction and valence bands) at zero energy energy for the retarded Green's function within the Keldysh model, Eq.~\ref{G}, and for the Green's function in the conventional SCBA. The latter is obtained using Eqs.~(\ref{eq:solutionXY}) as $G^R(\epsilon=0,\pp)=(X^B-Y^B{\bm \sigma}{\bm n}_{\pp})^{-1}$.
\begin{figure}
  \centering
  \includegraphics[width=3in]{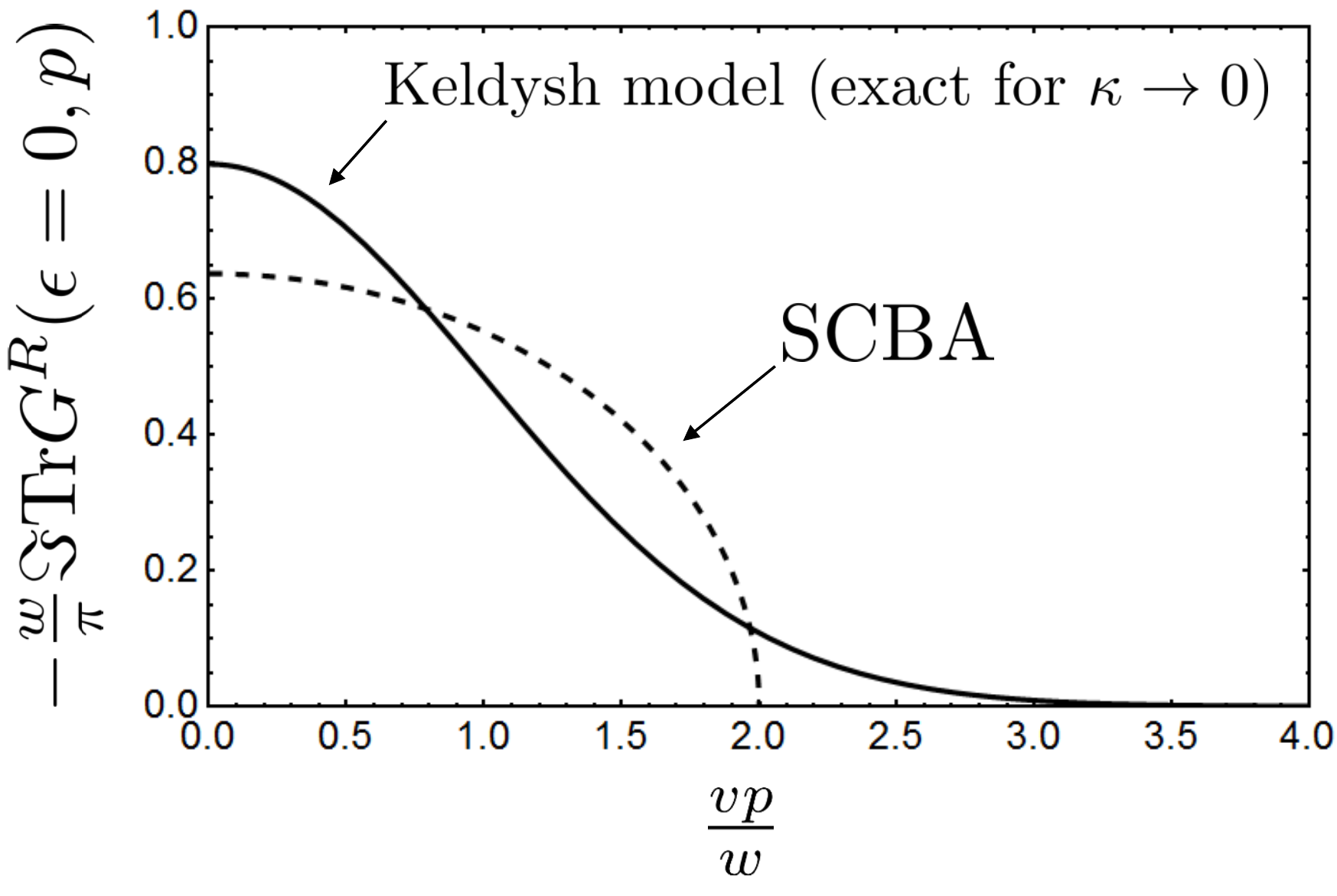}
  \caption{Full spectral weight at zero energy in the $\kappa\to0,\,w=\text{const}$ limit: Solid line - Keldysh model, dashed line - SCBA with full disorder correlator.}\label{fig:spectralweight}
\end{figure}

\subsection{Leading correction to the semiclassical density of states}

In this Section we calculate the leading order correction to the semiclassical DoS, Eq.~(\ref{nu}). To this end we use Eqs.~(\ref{eq:XY}) to find the first perturbative correction to $X_p$ and $Y_p$, that is, restric ourselves to the first Born approximation. It is sufficient to find the leading $O(v\kappa/w$) term in the DoS. To zero order in $w$ we take $X_p=\xi_+$ and $Y_p=vp$ and seek the first order correction by iterations:
\begin{eqnarray}
&&\hskip-.75cm
\delta X_p=-\frac{\kappa w^2}{\pi}\int^{+\infty}_{-\infty}\frac{\xi_+K_0(p,q)dq}{\xi^2_+-(vq)^2}+w^2\frac{\xi_+}{\xi^2_+-(vp)^2},\\
&&\hskip-.75cm
\delta Y_p=\frac{\kappa w^2}{\pi}
\int^{+\infty}_{-\infty}\frac{vqK_1(p,q)dq}{\xi^2_+-(vq)^2}-w^2\frac{vp}{\xi^2_+-(vp)^2}.
\end{eqnarray}
After performing the integrations and after some further algebra we find
\begin{eqnarray}
&&\hskip-.45cm
\delta X_p=w^2\left[\frac{\xi_+}{\xi^2_+-(vp)^2}-\frac{\xi_+}{(\xi_++iv\kappa)^2-(vp)^2}\right],\\
&&\hskip-.45cm\delta Y_p=\frac{w^2}{vp}\left[\frac{(vp)^2+(v\kappa)^2-iv\kappa\xi_+}{(\xi_++iv\kappa)^2-(vp)^2}-\frac{(vp)^2}{\xi^2_+-(vp)^2}\right]\nonumber\\
&&+\frac{iw^2v\kappa}{2(vp)^2}\ln\left(\frac{\xi_++iv\kappa+vp}{\xi_++iv\kappa-vp}\right).\end{eqnarray}
These expressions appear very singular, but one must bear in mind that it is only the full disorder-averaged Green's function that determines the DoS, and all singularities of the present perturbation theory are removed by Gaussian integration over $u$ in the expression for the retarded Green's function, Eq.~(\ref{Gfull}), expanded in the usual series in powers of $\Sigma$.

Knowledge of $\delta X_p$ and $\delta Y_p$ allows one to find correction to the density of states. We are only interested in the leading $\delta\nu\sim O(\kappa v/w)$ correction. Explicit calculation shows that the only term contributing to the DoS correction is $\delta\nu\propto\Im \sum_p g_0g_1\delta Y_p$. Following the standard path (expanding the Green's function to linear order in $\kappa$, taking its imaginary part, and performing the final momentum integrations), one finds
\begin{equation}
\delta\nu(\varepsilon)= \frac{w^2}{4\pi^2 v^3}\Im\frac{v\kappa}{\varepsilon-u+i0}.
\end{equation}
Finally, averaging this result with $F[u]$ from Eq.~(\ref{P}) we arrive at
\begin{equation}\label{nufinal}
\frac{\delta\nu(\varepsilon)}{\nu(0)}=-\sqrt{\frac{\pi}{8}}\frac{v\kappa}{w}\exp(-\varepsilon^2/2w^2).
\end{equation}
Thus the first perturbative correction to the DoS is regular and controlled by the smallness of $v\kappa/w$, as expected from the preceding discussion. We note that at no point did we have to introduce an artificial momentum cut-off, as is commonly done while treating the system in the self-consistent Born approximation. Taking into account the second Born approximation, or the first interference correction to the self-energy all lead to DoS-corrections that involve at least one additional power of $v\kappa /w$ compared to Eq.~(\ref{nufinal}), and can be ignored in the current treatment.

In order to establish the limits of validity of our results we need to determine the typical value of $w$. This can be done with the help of self-consistent Thomas-Fermi approximation. First recall that $\kappa=\sqrt{4\pi e^2\nu/\epsilon}$ and $w^2=2\pi ne^4/\epsilon^2\kappa$, which follows from Eq.~(\ref{uu}). Using now Eq.~(\ref{nu}) for the density of states we find an equation that defines the typical magnitude of the disorder potential fluctuation as a function of density $n$ in the form
\begin{equation}
w^4(\varepsilon^2+w^2)=\frac{2\pi^3}{g}\alpha^3n^2v^6,
\end{equation}
where $\alpha=e^2/v\epsilon$ is the effective interaction parameter. The most interesting case for us is the vicinity of the ``neutrality'' point where $\varepsilon\ll w$, so that
\begin{equation}\label{w}
w\simeq v\alpha^{1/2}n^{1/3},
\end{equation}
which sets the typical scale of the density of states and inverse screening length, respectively,
\begin{equation}\label{nuandkappa}
\nu\simeq\alpha n^{2/3}/v,\quad \kappa\simeq\alpha n^{1/3}.
\end{equation}
As discussed above, the Gaussian model of disorder amounts to a condition $n\kappa^{-3}\sim1/\alpha^3\gg1$. Perturbative treatment of the short-range disorder requires $v\kappa/w\sim\sqrt{\alpha}\ll1$. Both of these conditions can be satisfied provided that $\alpha<1$. Finally, we discuss validity of the semiclassical approximation that requires $|d\lambda/dx|\ll1$. The latter can be estimated as follows, $d\lambda/dx\sim(v/u^2)(du/dx)\sim(v/w^2)(w/\kappa^{-1})\sim\sqrt{\alpha}\ll1$.

\subsection{DoS in weak magnetic fields}

This semiclassical formalism can be readily extended to the case of finite magnetic field. The generic analysis of the density of states is quite involved. Here we will be primarily interested in the regime of weak fields, when magnetic length $l_B=\sqrt{1/eB}$ is large compared to the correlation radius of the disorder potential $l_B\gg\kappa^{-1}$. In this case we can use the same approach we have used for the magnetic-field-free case. In the strong-field limit, $l_B\ll\kappa^{-1}$, the treatment of Ref.~\onlinecite{Raikh} should be used. (See also Ref.~\onlinecite{Song2015} for an alternative quasiclassical approach to transport in the strong-field limit.)

The spectrum of Landau levels for the three-dimensional Weyl equation has the form
\begin{eqnarray}
  \varepsilon_m(p_z)&=& \pm\sqrt{\frac{2v^2}{l^2_B}|m|+(vp_z)^2},\quad m=1,2\ldots \\
  \varepsilon_0(p_z) &=& vp_z,
\end{eqnarray}
where we chose a particular chirality for the zeroth Landau level without loss of generality. We concentrate on the lowest energy limit where
\begin{equation}
\nu(0)=\frac{g}{2\pi l^2_B}\int^{+\infty}_{-\infty}\frac{dp_z}{2\pi}\sum^{+\infty}_{m=-\infty}\delta(\varepsilon_m(p_z)-u).
\end{equation}
Averaging this expression over $u$ with the help of Eq.~(\ref{P}), followed by the Gaussian integration over momentum $p_z$ and the geometrical series summation over $m$ one finds
\begin{equation}\label{nu-B}
\nu(0)=\frac{g}{(2\pi l_B)^2}\frac{1}{v}\coth\left(\frac{v^2}{2l^2_Bw^2}\right).
\end{equation}
Expanding this result in the low field limit one recovers Eq.~(\ref{nu}) to the lowest order with a \emph{positive} correction term $\delta\nu/\nu(0)=(v/l_B)^4/12w^4\propto B^2$. This finite field correction ultimately renders the change in the screening radius of the disorder potential, which in the case of Coulomb impurities will have an important effect on magnetoconductivity.

\section{Transport}

In this section we briefly discuss what self-consistent theory of screening in WSMs gives for the low-temperature transport, namely we concentrate on the regime of $T\ll vn^{1/3}$. In the semiclassical limit, provided that the mean free path $l$ for electron scattering is large, $k_Fl\gg1$, where $k_F$ is the Fermi momentum, the conductivity can be calculated from the Drude formula with energy dependent scattering time $\tau(\varepsilon)$, density of states $\nu(\varepsilon)$, and cyclotron frequency $\omega_c(\varepsilon)$,
\begin{equation}\label{sigma}
\sigma(B)=\frac{e^2}{6\pi}\int\frac{d\varepsilon}{4T\cosh^2\left(\frac{\varepsilon-u}{2T}\right)}\frac{v^2\nu(\varepsilon)\tau(\varepsilon)}{1+\omega^2_c(\varepsilon)\tau^2(\varepsilon)}.
\end{equation}
The ensemble-averaged transport scattering time for Coulomb impurities within the Boltzmann approximation can be calculated using the full disorder correlator, Eq.~(\ref{eq:fulldisorder}), since the $\delta(\qq)$ part of $D(\qq)$ from Eq.~(\ref{impurityline}) cannot lead to momentum transfer. The transport scattering time is thus given by
\begin{eqnarray}\label{tau}
&&\tau^{-1}=\frac{\pi\nu n}{2}\int^{\pi}_{0}d\theta|\phi(2k_F\sin\theta/2)|^2\sin\theta(1-\cos^2\theta)\nonumber\\
&&=\frac{\pi^3\alpha^2\nu nv^2}{k^4_F} [(2a^2+1)\ln(1+a^{-2})-2],\label{tau}
\end{eqnarray}
where $a=\kappa/2k_F$ and $\phi(q)$ is the Fourier transform of Eq.~(\ref{u}) taken at momentum $q=2k_F\sin\theta/2$, whereas the semiclassical expression for the cyclotron frequency for the linear spectrum is
\begin{equation}
\omega_c=v^2/(l^2_B\varepsilon).
\end{equation}
According to our earlier estimates of the screening radius one immediately concludes that $\kappa/k_F\sim\sqrt\alpha\ll1$, so it is sufficient to retain only the large logarithmic factor in Eq.~(\ref{tau}). Then, the zero field conductivity is estimated (up to an overall numerical factor) as follows
\begin{equation}\label{sigma-B-0}
\sigma\simeq e^2v^2\nu(u)\tau(u)\simeq \frac{e^2n^{1/3}}{\ln(1/\alpha)},
\end{equation}
where we took the typical value of potential $u\sim w$ and used Eq.~(\ref{w}). One can check that derived expression indeed corresponds to the Boltzmann limit. We note in passing that a local conductivity tensor is appropriate for the situation considered here, since Eqs.~(\ref{w}), (\ref{nuandkappa}), and (\ref{tau}) imply that $v\tau\kappa\sim1/|\ln\alpha|\ll 1$; in other words, the transport mean free path is much shorter than the screening length.  Further, it is worth mentioning that the dependence of $\sigma$ on the interaction parameter $\alpha$ was a subject of controversy since different results exist in the literature.

The field dependence of the conductivity comes from two factors. The first is obviously due to the cyclotron motion, which we estimate by expanding the denominator of Eq.~(\ref{sigma})
\begin{equation}\label{sigma-B-a}
\frac{\delta\sigma(B)}{\sigma}\simeq -[w_c(u)\tau(u)]^2\simeq-\frac{1}{\alpha^7\ln^2(1/\alpha)}\frac{1}{l^4_Bn^{4/3}}.
\end{equation}
Curiously, there is another contribution, of the opposite sign, that stems from the already mentioned fact, that magnetic field modifies the screening radius. Since the product $\nu\tau$ is inversely proportional to $\ln(k_F/\kappa)$ and $\kappa\propto\sqrt{\nu}$ then it follows from the low field asymptotic of Eq.~(\ref{nu-B}) that for this mechanism
\begin{equation}\label{sigma-B-b}
\frac{\delta\sigma(B)}{\sigma}\simeq\frac{\delta\kappa(u)}{\kappa\ln(1/\alpha)}\simeq\frac{1}{\alpha^2\ln(1/\alpha)}
\frac{1}{l^4_Bn^{4/3}}.
\end{equation}
For weak interactions, $\alpha\ll1$, Eq.~(\ref{sigma-B-a}) always dominates by a parametrically large factor and leads to positive magnetoresistance at weak field that eventually crosses over to quantum regime at higher fields, $\sigma(B)\sim (e^2/v)nl^2_B$. \cite{Abrikosov,Klier} However, for some material systems, where interaction is moderately strong, $\alpha\sim1$, one in principle could envision a scenario where the fate of the sign of magnetoconductivity will be determined by the overall numerical prefactors in Eqs.~(\ref{sigma-B-a}) and (\ref{sigma-B-b}). Obviously the criterion $\alpha\lesssim1$ limits the applicability of our estimates.

An elaborate theory of magnetoresistance in WSMs was recently developed in Ref.~\onlinecite{Klier} in various regimes determined by the relation between the relevant energy scales $T$, $v/l_B$, and $vn^{1/3}$. Our estimates for $\delta\sigma(B)$ at weak field are consistent with that of Ref.~\onlinecite{Klier} with an addition of the contribution governed by Eq.~(\ref{sigma-B-b}), where in Eqs.~(\ref{sigma-B-0})--(\ref{sigma-B-b}) we also retained functional dependence on the interaction parameter $\alpha$.

\section{Summary}

To conclude, we have studied the density of states and transport properties of Weyls semimetals with the long-range Coulomb disorder. Provided that the semiclassical limit is satisfied, DoS induced by the long-range disorder potential fluctuations can be found exactly. In addition, we have developed systematic perturbation theory to account for corrections due to the short-range scattering. Concerning the transport properties, for weak interactions and magnetic field the magnetoresistance is expected to be positive for a single Weyl node. (As compared to a negative magnetoresitance for current flow along the magnetic field when the realistic multivalley situation is considered. \cite{Son-Spivak}) However, based on the presented analysis, one should not exclude a possibility of nonmonotonic magnetoresistance in systems with relatively strong interactions.

\begin{acknowledgements}
We would like to thank I.~Gornyi, J.~Klier, A.~Mirlin, and M.~Raikh for fruitful discussions. We are grateful to M. Dzero for reading and commenting on the paper. This work was supported by NSF Grant No. DMR-1409089 (D.A.P.), by NSF Grant DMR-1401908 and in part by DAAD grant from German Academic Exchange Services, and SPP 1666 of the Deutsche Forschungsgemeinschaft (A.L.), and DOE Basic Energy Sciences Grant No. DE-FG02-06ER46313 (E.G.M.). D.A.P and A.L. acknowledge the hospitality of Spin Phenomena Interdisciplinary Center at Johannes Gutenberg-Universit\"{a}t Mainz, where this work was finalized.
\end{acknowledgements}

\end{document}